\documentstyle[floats,preprint,aps,epsf]{revtex}

\draft
 
\newcommand{\be}{\begin{equation}}
\newcommand{\ee}{\end{equation}}
\newcommand{\ba}{\begin{eqnarray}}
\newcommand{\ea}{\end{eqnarray}}

\begin{document}

\title{ Chaos and Rotating Black Holes with Halos}

\author{ P.S. Letelier\footnote{e-mail: letelier@ime.unicamp.br}
 and 
W. M. Vieira\footnote{e-mail: vieira@ime.unicamp.br} } 
 
\address{
 Departamento de Matem\'atica Aplicada-IMECC,
Universidade Estadual de Campinas,
13081-970 Campinas,  S.P., Brazil}
\maketitle
 
\begin{abstract}
 
 The occurrence of chaos for  test particles 
moving  around a slowly rotating black hole  with 
a dipolar halo is studied using Poincar\'e sections. We find 
a novel effect, particles with angular momentum opposite to 
the black hole rotation have  
larger chaotic regions in phase space than  particles initially moving 
in the same direction.

\end{abstract}

The main  paradigm  in the study of the motion of stars in 
a galaxy is a model of a central bulge surrounded by a 
halo \cite{bin}. In this context, for the particular
 axially symmetric case, arose the 
 celebrated  H\'enon-Heiles model  \cite{HH} whose study has
 been the  source of inspiration 
 of many  researches  on chaotic behavior \cite{guho}. 
 The underlying theory in
 this case is 
the usual Newtonian Gravitation that for large masses and velocities 
is known to be less appropriate than  Einsteinian  General Relativity. 
In the late case the Newtonian potential is replaced by the 
spacetime metric and  Newton motion equations by geodesics. 
This change of dynamics can produce dramatic effects, for
 instance,  test particles moving in the presence of
 systems of masses that are integrable in Newtonian theory are chaotic
in General Relativity,  examples are: the fixed two body
  problem  \cite{cont,ssmae},
and particles moving in a monopolar 
center of attraction surrounded by a dipolar halo \cite{wldi}.
Also  gravitational waves, a non existing phenomenon in the
 Newtonian realm, can produce irregular motion of test particles
 orbiting around a static black hole \cite{bc,lwmelgr}.
Another  distinctive feature of general relativity is the dragging 
of inertial frames due to mass rotation. This fact is
  observed, for instance,  in the impressive
differences of the geodesic motion in Schwarzschild
 and Kerr geometries \cite{chandra}. 

In this Letter we study the effects of rotation in the motion
 of a particle orbitating around a slowly rotating  black
 hole surrounded by a dipolar halo.
The non rotating case is chaotic, so we shall study mainly the
 change in the chaotic behavior due to the rotation of the
 center of attraction.  A typical situation is represented by a galaxy
with a rapid rotating center surrounded by a distant massive halo, ring or other
 shell-like  distributions of matter.  We study the motion of particles 
moving between the center and the  halo whose first  contribution
is dipolar. This  contribution is always  present whenever the halo does not possess
reflection symmetry with respect to the black hole equatorial plane.

The metric that represents the superposition of a Kerr black hole and
  a dipole along the rotation axis is a stationary axially symmetric
 spacetime. The vacuum Einstein equations for this class of
  spacetimes is an 
integrable system of equations 
that is closely related to the principal sigma model
 \cite{cosgrove}. Techniques to actually find the solutions
 are B\"acklund 
transformations and the inverse scattering method, also 
 a third method  constructed with elements of the previous
 two is the ``vesture method", all these methods are 
closely related \cite{cosgrove}.
 The general metric that represents the nonlinear superposition of
a Kerr solution with a Weyl solution, in particular,
 with a multipolar expansion can be found by  using the ``inverse scattering 
  method"   \cite{sol}.   We find
\be 
ds^2=g_{tt}(r,z)dt^2+2 g_{t\phi}(r,z)dt d\phi+g_{\phi\phi}(r,z)d\phi^2+ f(r,z)(dz^2+dr^2),
\label{m1}
\ee  
where
\begin{eqnarray}
&&g_{tt}=-[e^{-2{\it D} p u v}(e^{-4{\it D} p u} (e^{-8{\it D} p v} (p  +1)^{2} (u^2 -1)  +e^{-8{\it D} p}  (p -1)^{2}(u^2  -1 )\nonumber\\
&&\hspace{2cm}-2 e^{-4{\it D} p(v+1)}(u^2-v^2 )q^{2})+e^{-4{\it D} p(1+v)}(e^{-8{\it D} pu}+1)(v^2 -1) q^{2})]/F ,\nonumber\\
 &&g_{t\varphi}=2 p qe^{-2{\it D} p (u+v+1)} [e^{-4{\it D} p u} (e^{-4 {\it D} p v} (p+1 ) (u-v)(u +1 )\nonumber\\
&& \hspace{4cm} +e^{-4{\it D} p}(p -1) (u +v )(u -1 ))(v+1)) \nonumber\\
&&\hspace{2cm}-e^{-4{\it D} p v} (p +1)(u +v )(u+1)(v-1)
 -e^{-4 {\it D} p}(p-1)(u-v )(u-1)(v-1)]/F , \nonumber \\
&& g_{\varphi \varphi}=[(g_{t\varphi})^2-(mr)^2]/g_{tt} ,\nonumber\\
&& f= m^2e^{-4{\it D} p((v+1)u-v+1)} [e^{-4 {\it D} p u v}(p-1)^{2}(u-1)^{2}+e^{-4{\it D}p((v+1)u+v-1)}(v+1)^{2}q^{2}]\nonumber\\
&& \hspace{4cm}+e^{-4{\it D} p((v+1)u+v-1) -4{\it D} p u v} (p +1)^{2} (u+1) ^{2}\nonumber \\
&&  \hspace{2cm}-2 e^{-2 {\it D} p((v+1) u +v-1)-2 {\it D} p((v+1)u -v +1) -4{\it D} p u v}(u^2-v^2)q^{2}+e^{-8{\it D} p u v} (v -1)^{2} q^{2})/H ,\nonumber \\
&& F\equiv e^{-4{\it D} p u} [e^{-8{\it D} p v} (p +1 ) ^{2}(u+1 ) ^{2}+e^{-8{\it D} p} (p-1)^{2} (u-1)^{2}-2 e^{-4 ){\it D} p(v+1} (u^2 -v^2) q^{2}] ,\nonumber\\
 &&H\equiv 4 \exp[-2 {\it D} p((v +1) u +v-1) -2{\it D} p((v+1)u-v+1 ) \nonumber  \\
 &&\hspace{6cm}-6{\it D} p uv-(u^2-1)(v^2-1){\it D}^{2}p^{2}] .
\label{BZ}
\end{eqnarray}
 The coordinates $(r,\phi, z)$ are dimensionless and have the range of the usual cylindrical 
coordinates. They are related to $u$ and $v$ by: $z=uv$ and $r=(u^2-1)^{1/2}(1-v^2)^{1/2}$, $u\geq 1$ and $-1\leq v \leq -1$. Our units are such that  $c=G=1$; {\it D} represents the dipole
 strength and $a$ the rotation parameter, $q=a/m$ and $p^2+q^2=1$. The coordinate transformation
$ t^\prime=t+2a\varphi, u=R/m-1, v=\cos\vartheta, \varphi^\prime=\varphi$
 reduces (\ref{BZ}) with $D=0$ to the  Kerr solution in the usual Boyer-Lindquist coordinates \cite{BZ}.

To study the slow rotation case is better to  use the 
metric  obtained by keeping the   first 
order terms in  the rotation 
parameter $a$ in the exact metric (\ref{BZ}).  This approximation, for the parameters and
 range of coordinates used, will not produce a
 significant information loss; we shall comeback to this point later. We find for $g_{\mu\nu}=g_{\mu\nu}^0 + a g_{\mu\nu}^1 $,
 \ba
g_{tt}&=&-\frac{u-1}{u+1}\exp(-2{\it D}uv), \nonumber\\
g_{t\phi}&=&\frac{a}{u+1}[(u+v)(1-v)\exp(-2{\it D}(1-u-v)), \nonumber\\
&&\hspace*{1cm}+(u-v)(1+v)\exp(-2{\it D}(1+u-v))],\nonumber\\
g_{\phi\phi}&=&m^2(1-v^2)(1+u)^2\exp(2{\it D}uv), \nonumber\\
f&=&m^2\frac{(u+1)^2}{u^2-v^2} \exp ({\it D}[(u^2-1)(v^2-1){\it D}+2uv-4v+4]).
\label{LT}
\ea 
 
The geodesic equations for the metric (\ref{m1}) can be cast as
\ba
&&\dot{t}=g^{tb}E_b,\hspace{0.5cm} \dot{\phi}=g^{\phi b}E_b , \label{geo1} \\
&&\ddot{r}=-\frac{1}{2f}[g^{ab}_{,r}E_a E_b +f_{,r}(\dot{r}^2
-\dot{z}^2)+2f_{,z}\dot{r}\dot{z}], \label{geo2}\\
&&\ddot{z}=-\frac{1}{2f}[g^{ab}_{,z}E_a E_b +f_{,z}(\dot{z}^2
-\dot{r}^2)+2f_{,r}\dot{r}\dot{z}], \label{geo3}
\ea 
where the dots denote derivation with respect to $s$ and the indices
$a$ and $b$  take the values $(t,\phi)$, $g^{ab}$ stands for 
 the inverse of $g_{ab}$. $E_t=-E$ and $E_\phi=L$ are integration 
constants; $E$ and $L$ are the test  particle energy and angular momentum,
 respectively.
The set (\ref{geo1})--(\ref{geo3}) admits a third integration constant
\be
E_3=g^{ab}E_a E_b+f(\dot{r}^2+\dot{z}^2)=-1 . \label{e3}
\ee %
Thus to have complete integrability we need one more independent constant
 of integration. 
In the case of pure Kerr solution (${\it D}=0$) we have a fourth
constant due to the existence of a Killing tensor and for
the non rotating case we have another constant related to a third
Killing vector associated to spherical symmetry \cite{chandra}.

 The  system   (\ref{geo2})--(\ref{geo3}) can be written as a four
 dimensional dynamical system in the variables $(r, z, P_r= 
\dot{r}, P_z=\dot{z})$. A  convenient method to study 
qualitative aspects of this system
  is to compute the Poincar\'e sections through  the plane
  $z=0$. The intersection  of the orbits with this plane will be
 studied in some  detail for bounded motions. We shall
 numerically solve the system  (\ref{geo1})--(\ref{geo3}) and use 
the integral (\ref{e3})
to control the accumulated error along the integration; we shall return to 
this point later.

 The Poincar\'e section for different
initial conditions    with 
  energy $E=0.965$ and angular momentum $L=-3.75$   (counter rotation) 
moving in an approximate Kerr
geometry (${\it D}=0$) with rotation parameter $a=0.01$ and mass $m=1$
(this value for the mass will be kept unchanged from now on) are presented 
in Fig. 1.  We have the typical section of an integrable motion, i.e., 
the sectioning   of invariant tori, for integrability and KAM
 theory, see for instance \cite{arnIII}.

 We also studied the same  case  for direct rotation $L=3.75$, as well as, the 
corresponding Schwarzschild limit $a=0$, and $L=-3.75$.    All 
these  cases present Poincar\'e sections almost identical to
 Fig. 1. The section area for  the Schwarzschild case is slightly 
smaller than the section area of Fig. 1  and  the section area
for particles in direct rotation in a Kerr geometry is even smaller. 
We have that counter rotation enlarges the area of the section and direct 
rotation shrinks it.  This is clearly an effect of the dragging of inertial frames
due to rotation.

The motion of test particles around 
 a static black hole with a  dipolar halo ($a=0$ and ${\it D}\not=0$ ) is
 chaotic and it was studied in some detail in  \cite{wldi} for
a different energy shell.  In Fig. 2 we show
the Poincar\'e section for ${\it D}=0.0005$ and the same values of
  $E=0.965$ and $L=3.75$ as in Fig. 1. We find  islands of integrability 
surrounded 
by chaotic motion. The two isolated islands
around the points (10, 0.05) and (5, 0.075) are parts of the same
 torus. In the case studied in \cite{wldi}  they were closer.

 Now we shall consider  a  particle moving around  an slowly rotating 
attractive 
center surrounded by a  dipolar halo for both direct rotation
 and counter rotation. In Fig. 3 we draw the Poincar\'e section 
for   
${\it D}=0.0005$, $a=0.01$,
$E=0.965$ and $L=3.75$ (direct rotation). We see that the islands
of stability are larger in this case than in the non rotating
 case (see Fig. 2);
also we have new systems of small islands immersed in the 
chaotic region.  We
 have that the chaotic region is smaller
in this case than in the equivalent non rotating one. It does 
look like that the direct rotation diminishes the effect of the dipolar 
strength as a chaos source. In Fig. 4 we present the
 section with the same parameters
 of Fig. 3, except that now we have counter  rotation, $L=-3.75$. In 
Fig. 4 we observe that  the chaotic region  increases in a significant
 way and also that the  islands 
located around the points (10, 0.05) and (5, 0.075) in Fig. 2 and 3
 has disappeared in this scale. In other words, with $D\not = 0$, 
 the counter rotating 
motion of  particles is  more chaotic than the static case, the later 
being  more chaotic than the direct rotation case. Also, we can 
say that
the counter (direct) rotation reinforces (weakens) the strength of 
the dipole as a  source
of the chaotic motion  (for  ${\it D}=0$ we have an
  integrable system). This effect is a manifestation of the fact that
particles moving in non equatorial orbits of a rotating central body 
can suffer  repulsive  forces due to rotation. This fact  allows
in a Kerr geometry closed orbits of test  particles moving on planes 
parallel to the equatorial plane
 \cite{bonnor}, though these orbits are not stable.
Although in the present Letter we present results 
for particular values of the parameters involved, we did a rather 
extended  numerical study that supports  our conclusions, Figs. 3, and 4
  being representative of this search. 

 For the values of the parameters  ${\it D}=0.0005 $, $a=0.01$,
$E=0.965$ and $L=\pm 3.75$, we have that the particles move in the ``box"
$4.7<r<21$, $-5< z <9$. We take as a  measure of error the quantities 
\be
\Delta g_{ab}=|(g_{ab}^{ex}-g_{ab})/g_{ab}^{ex}|,  \hspace{0.5cm}
\Delta f= |(f^{ex}-f)/f^{ex}|, \label{del}
\ee 
where  in  these expressions the   sum rule of repeated 
indices does not apply.   $g_{\mu\nu}^{ex}$  and $ g_{\mu\nu}$
 refer to the solutions (\ref{BZ}) and (\ref{LT}),  respectively.
We find that for the above mentioned range of coordinates 
and the values of parameters used in this Letter the quantities 
defined in (\ref{del}) are   at most of  the order
of $10^{-6}$. Also in this range,  the error in the derivatives of 
the metric functions  is even smaller (the metric functions are
 very smooth).  We also want to mention that  the Poincar\'e sections 
shown in this Letter were computed from
orbits with  an accumulated error in the ``energy" [cf. Eq. (\ref{e3})]
 smaller than  $10^{-10}$.

 We want to conclude with a discussion of a  possible astrophysical 
implication of  our main result: direct rotating particles are
 less chaotic than counter rotating ones.
In   a rotating  center of gravitational attraction  we can have 
structures formed by counter  and direct rotating  particles, thus
 our result favors larger life 
times of structures formed with direct rotating particles. A deeper
 discussion of this point will be presented elsewhere. 
  
The authors thank CNPq and FAPESP for support and
 to S.R. de  Oliveira and N. Santos  for discussions.


\newpage
\noindent
{\bf FIGURE CAPTIONS}

\vspace*{1cm}
\noindent
Fig. 1.    Poincar\'e section of  test 
particles moving with angular momentum $L=-3.75$  in an approximate 
 Kerr geometry with rotation parameter $a=0.01$ (counter rotation) and mass
 $m=1$.  This is a typical section of an integrable system. \\
 
\vspace*{0.4cm}
\noindent
Fig. 2.  Poincar\'e section for ${\it D}=0.0005$,
  $E=0.965$, $a=0$, $m=1$ and $L=\pm 3.75$.  The two isolated islands
around the points (10, 0.05) and (5, 0.075) are parts of the same
 torus.\\ 

\vspace*{0.4cm}
\noindent
Fig. 3.  Poincar\'e section for   ${\it D}=0.0005$, $a=0.01$, $m=1$
$E=0.965$ and $L=3.75$ (direct rotation). The islands
of stability are larger in this case than in the non 
  rotating case (cf. Fig. 2).\\

\vspace*{0.4cm}
\noindent
Fig. 4. Poincar\'e section with the same parameters
 of Fig. 3 , except that now $L=-3.75$ (counter  rotation). The chaotic 
region  increases in a significant way, also  the system of islands 
located around the points (10, 0.05) and (5, 0.075) in the two precedent
 cases  has disappeared.
\noindent

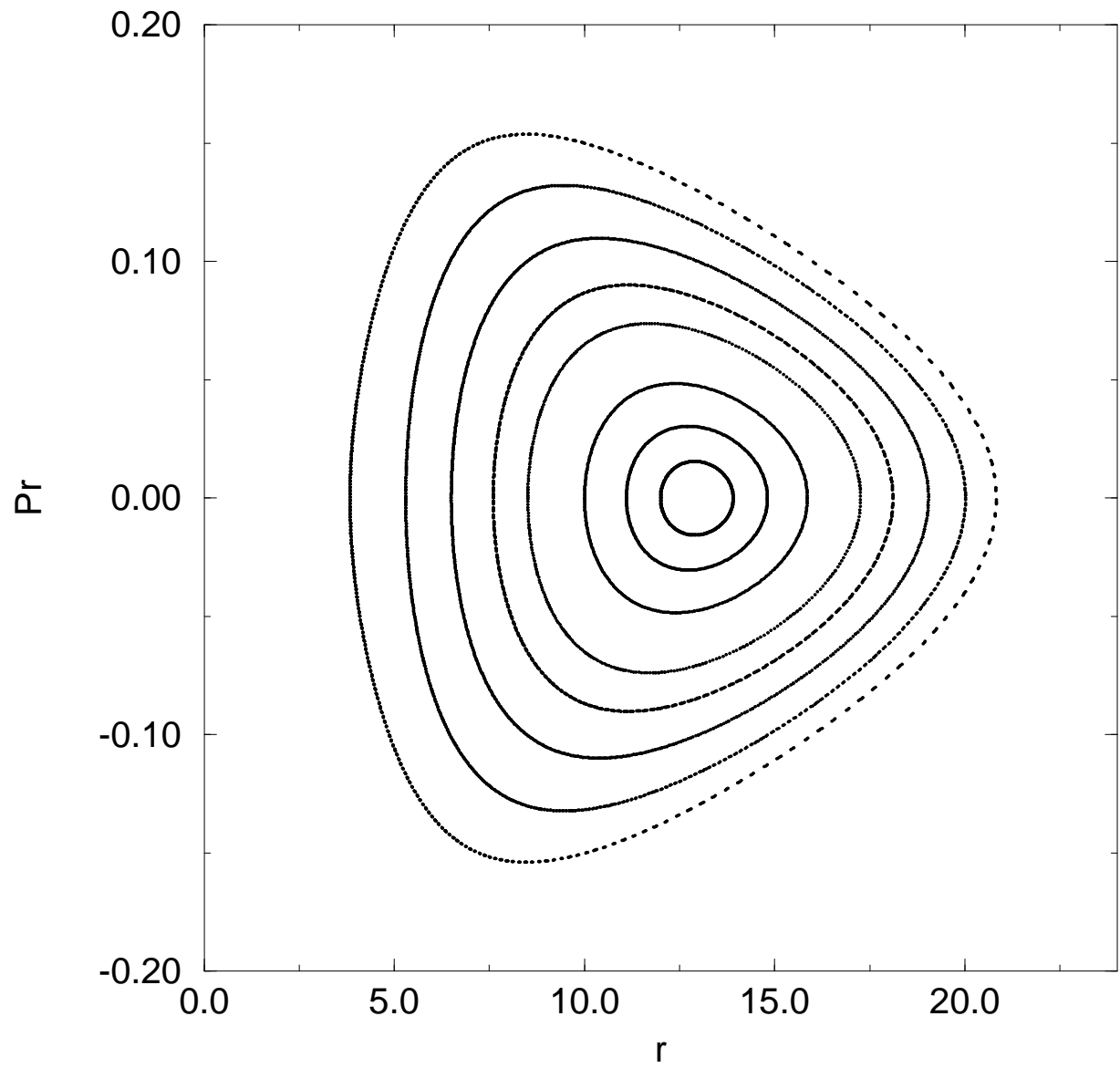
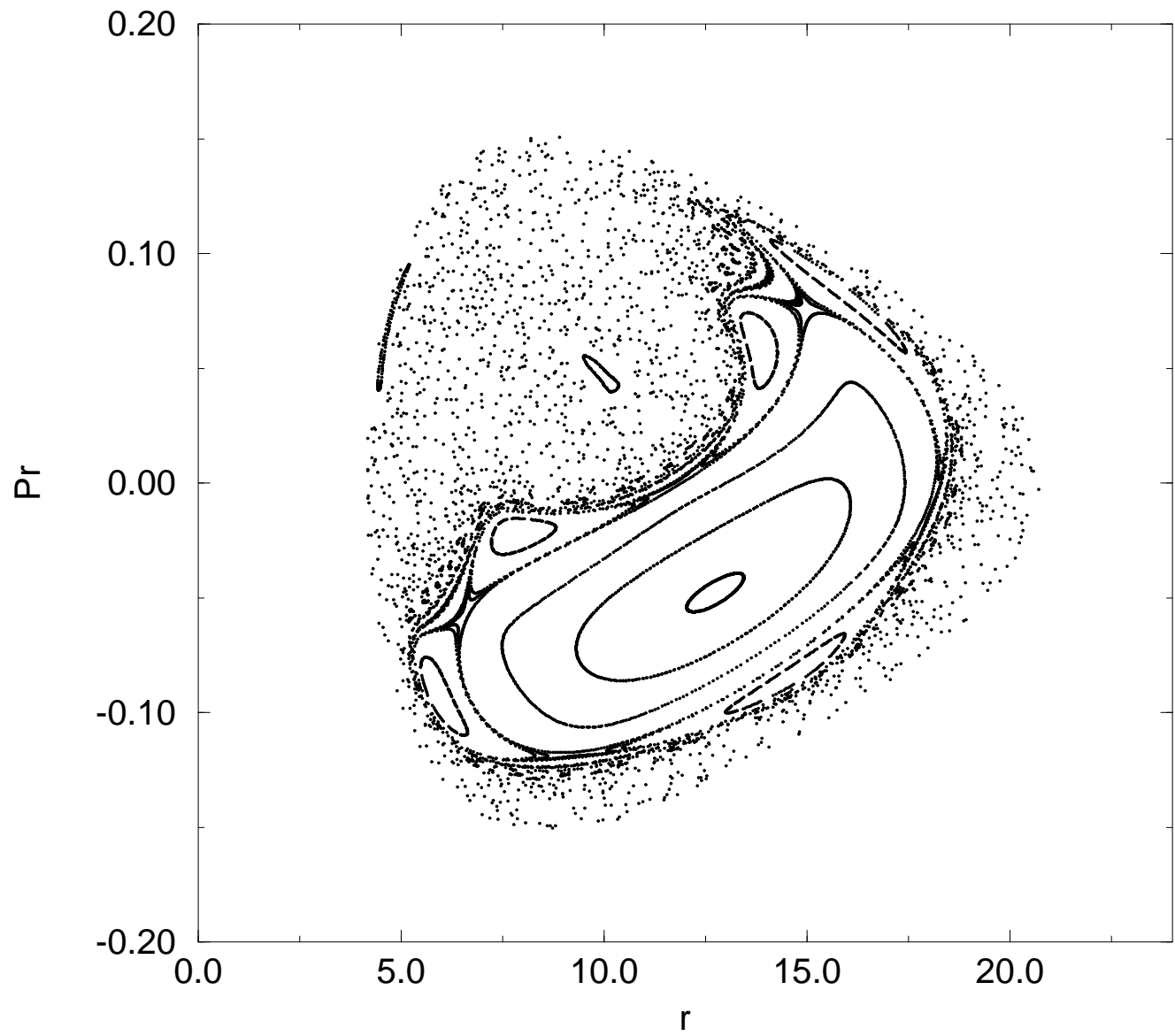
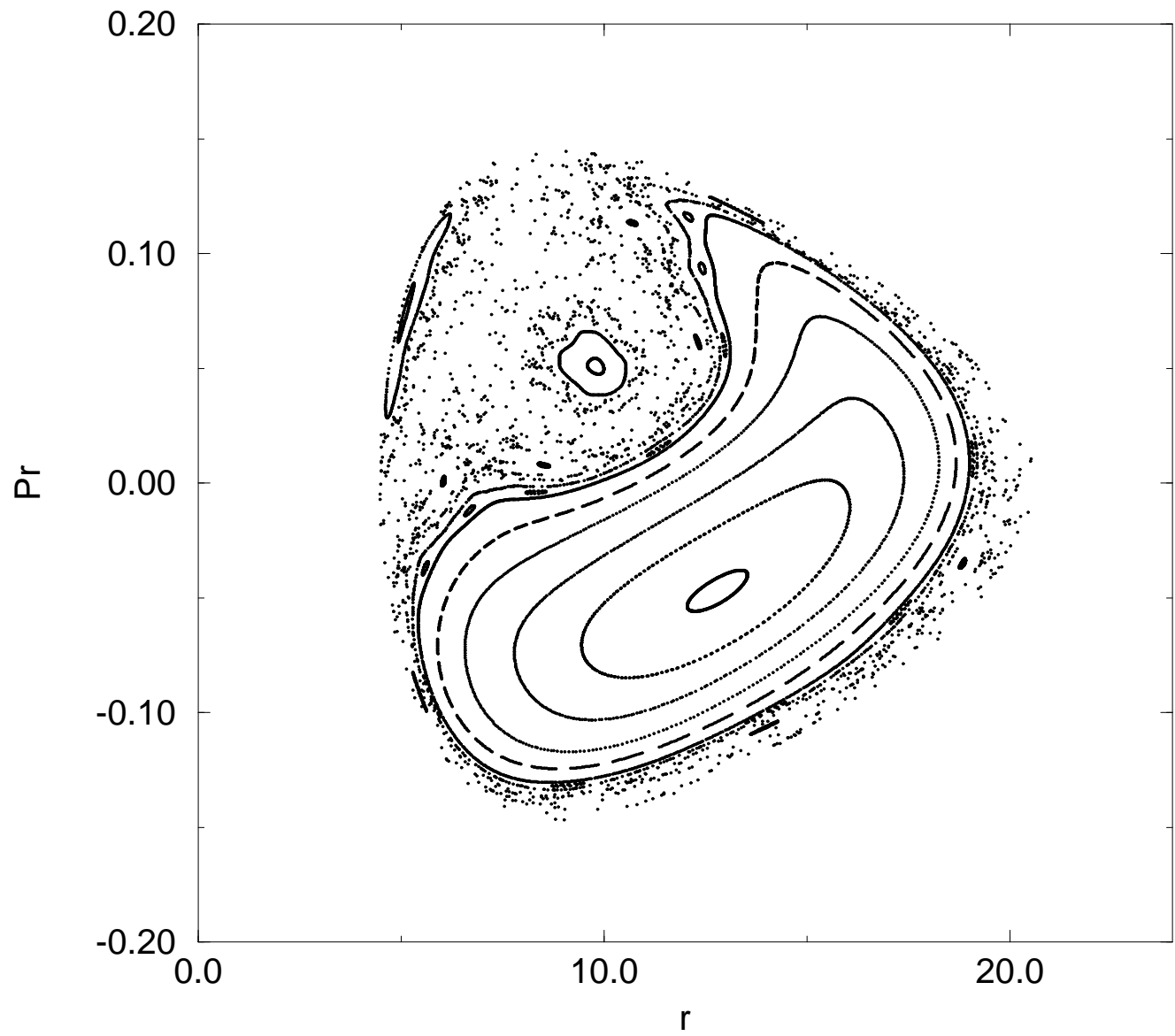
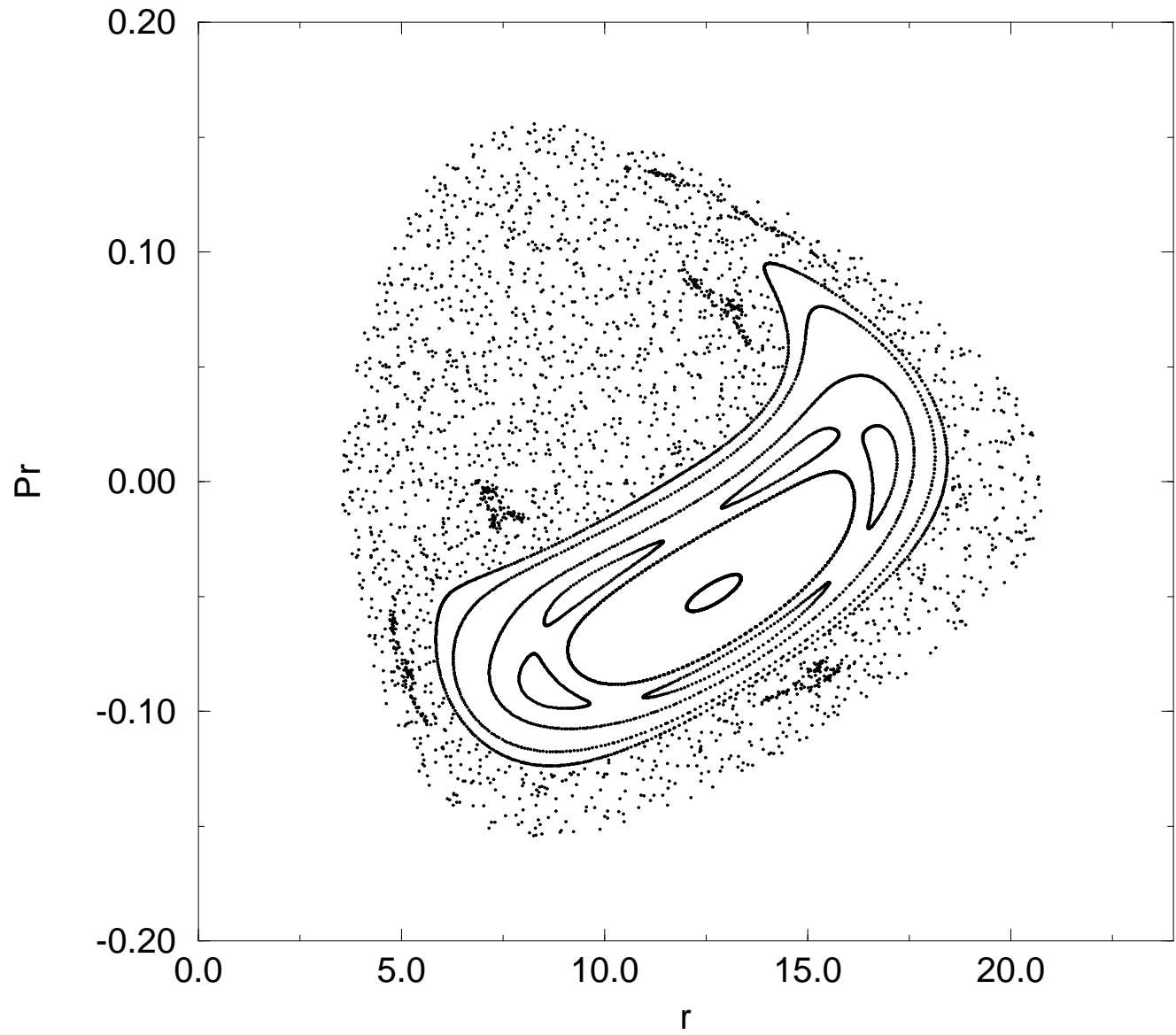
\end{document}